\documentclass[prb,superscriptaddress,showpacs]{revtex4-1}
\usepackage{graphicx}
\usepackage{dcolumn}
\usepackage{bm}
\usepackage{amsmath}
\usepackage{amssymb}
\usepackage{ulem}
\usepackage{color}
\usepackage{float}

\def\B{{\mathcal B}}
\def\NB{{N_\B^{}}}
\def\NL{{M_\Lambda^{}}}

\def\Tr{{\bf Tr}}
\providecommand{\includegraphics}[2][width=\textwidth]{$#2$}
\begin{document}

\title{The Hartree-Fock phase diagram of the two-dimensional electron gas}
\author{B. Bernu}
\affiliation{LPTMC, UMR 7600 of CNRS, Universit\'e P. et M. Curie, Paris, France}
\author{F. Delyon}
\affiliation{CPHT, UMR 7644 of CNRS, \'Ecole Polytechnique, Palaiseau, France}
\author{M. Holzmann}
\affiliation{LPTMC, UMR 7600 of CNRS, Universit\'e P. et M. Curie, Paris, France}
\affiliation{Univ. Grenoble 1/CNRS, LPMMC UMR 5493, Maison des Magist\`{e}res, 38042 Grenoble, France}
\author{L. Baguet}
\affiliation{LPTMC, UMR 7600 of CNRS, Universit\'e P. et M. Curie, Paris, France}

\date{\today}
\begin{abstract}
We calculate the ground state phase diagram of the homogeneous electron gas in two dimensions within the
Hartree-Fock approximation. At high density, we find  stable solutions, where the electronic charge and spin density 
form an incommensurate crystal having more crystal sites than electrons, whereas
the commensurate Wigner crystal is favored 
 at lower densities, $r_s \gtrsim 1.22$.  
 Our explicit calculations demonstrate that
 the homogeneous Fermi liquid state -- though being an exact  stationary solution of the Hartree-Fock equations --
 is never the Hartree-Fock ground state of the electron gas.
\end{abstract}
\maketitle

\section{Introduction}

Electrons are found everywhere in matter, most of the time localized by positive charges.
In typical condensed matter situations, electronic densities and temperatures are 
such that, in addition to the external positive charges,  a quantum description of electrons interacting with  each other
is necessary, leading in general to a difficult quantum many-body problem. 
The homogeneous electron gas, where the positive charges are reduced to solely ensure global electro-neutrality,
is one of the most fundamental model to study electronic correlation effects.  In three dimensions, $d=3$, 
valence electrons in alcaline metals realize the electron gas to high precision, in particular in solid sodium  \cite{Na},
whereas the two dimensional electron gas, $d=2$, (2DEG), and its extension to quasi-two dimensions\cite{quasi2D} is relevant for  electrons at
heterostructures, e.g. semiconductor-insulator interfaces\cite{hetero}. At zero temperature, the electron gas
is described by a single parameter, the density $n$ or equivalently by the dimensionless parameter $r_s=a/a_B$. Here $a = [2(d-1)\pi  n/d]^{-1/d}$ is the mean inter particle distance,
and $a_B=\hbar^2/(me^2)$ is the Bohr radius, where $-e$ and $m$ are the electronic charge and mass, respectively.

As pointed out by Wigner \cite{Wigner}, at low densities and zero temperature, 
electrons will form a crystal, supposed to melt at higher densities where the kinetic energy dominates over the
interaction. In the limit $r_s \to 0$, the Hartree-Fock approximation (HF) applies.
Since the non-interacting Fermi sea  remains a stationary solution of the Hartree-Fock equations, it is natural to assume a Fermi liquid phase at high densities. First principle calculations, such as Quantum Monte Carlo \cite{QMC3D,exchange3D,QMC2D,exchange},
have located the transition from the Wigner crystal (WC) to the homogeneous Fermi liquid (FL) to high precision.
Still,  there are indications that the Fermi liquid phase is not necessarily the absolute
ground state of the electron gas at high densities \cite{Overhauser, Giulani,HF-2008,Shiwei} and
that a direct transition between Wigner crystal and a homogeneous Fermi liquid cannot occur in two dimensions
in the thermodynamic limit\cite{Spivac, Hexatic,Waintal}.
These conjectures actually hold already for the electron gas in the Hartree-Fock approximation, but, despite  the
early predictions by Overhauser of the
spin and charge density instability of the Fermi liquid ground state, explicit, numerical HF calculations \cite{Needs} have 
not confirmed them for a long time. 
Based on Bloch functions, these HF calculations\cite{Needs} studied unpolarized and polarized Wigner crystal 
phases of square and triangular symmetries and found 
a first-order transition to the unpolarized Fermi gas which -- within this study -- remains the lowest energy state for $r_s\lesssim 1.44$.
Only recently, the first
self consistent Hartree-Fock solutions with energies below the Fermi liquid energy have been found at high densities \cite{HF-2008}.

The HF solutions of Ref.~\cite{HF-2008} obtained without imposing any periodicity  in the density
show that the fully polarized electron gas in two dimensions
forms a periodic charge density 
with triangular symmetry at high densities. In contrast to the low-density Wigner crystal, the 
number of maxima of the charge density is higher than the number of electrons, having thus metallic character, and we will
refer to such states as  incommensurate crystals in the following. However, incommensurate states give rise to important
size effects, and the calculations in Ref.~ \cite{HF-2008} were limited to $\sim 500$ electrons.

In this paper, 
we extend the desctiption based on Bloch waves to study arbitrary modulation and occupation number. We focus on 
the density region $r_s<4$, where incommensurate states may occur.
We show how the incommensurate states can be represented by the vector $Q$ of the charge modulation.
Restricting the search for the
HF ground state to states with arbitrary $Q$,  we are able to overcome size restrictions and we explore the 
phase diagram of the 2DEG including triangular ($\blacktriangle$) and square ($\blacksquare$) symmetries. 
While our minimization also includes the possibility of partial polarized states,  they do not occur as ground states which
are either unpolarized (U) or fully polarized (P); in particular,
we show that  the incommensurate unpolarized crystal is favored
at high densities. 
Whereas the momentum distribution of the Wigner crystal is a continuous
function of the momentum, we show that there are angle-selective steps in the incommensurate phase.

\section{Methods}
\label{SEC-P2DEG} 

The Hamiltonian of the electron gas containing $N$ electrons writes
\begin{eqnarray}
	H=-\frac{1}{2}\sum_i  \Delta_i+\sum_{1\leq i<j\leq N_p}v(x_i-x_j)
	\label{Hamiltonian}
\end{eqnarray}
where $\Delta_i$ is the Laplacian with respect to $x_i$,  $v(x)$ is the electrostatic interaction $v(x)=\|x\|^{-1}$, and we have used atomic units where distances are measured in units of $a_B$ and energies in Hartree, $1 \text{Ha}=\hbar^2/(ma_B^2)$. In addition to Eq.~(\ref{Hamiltonian}), the interaction between electrons and a positive background charge must be considered to ensure charge neutrality.

\begin{figure}
\begin{center}
\includegraphics[width=0.99\textwidth]{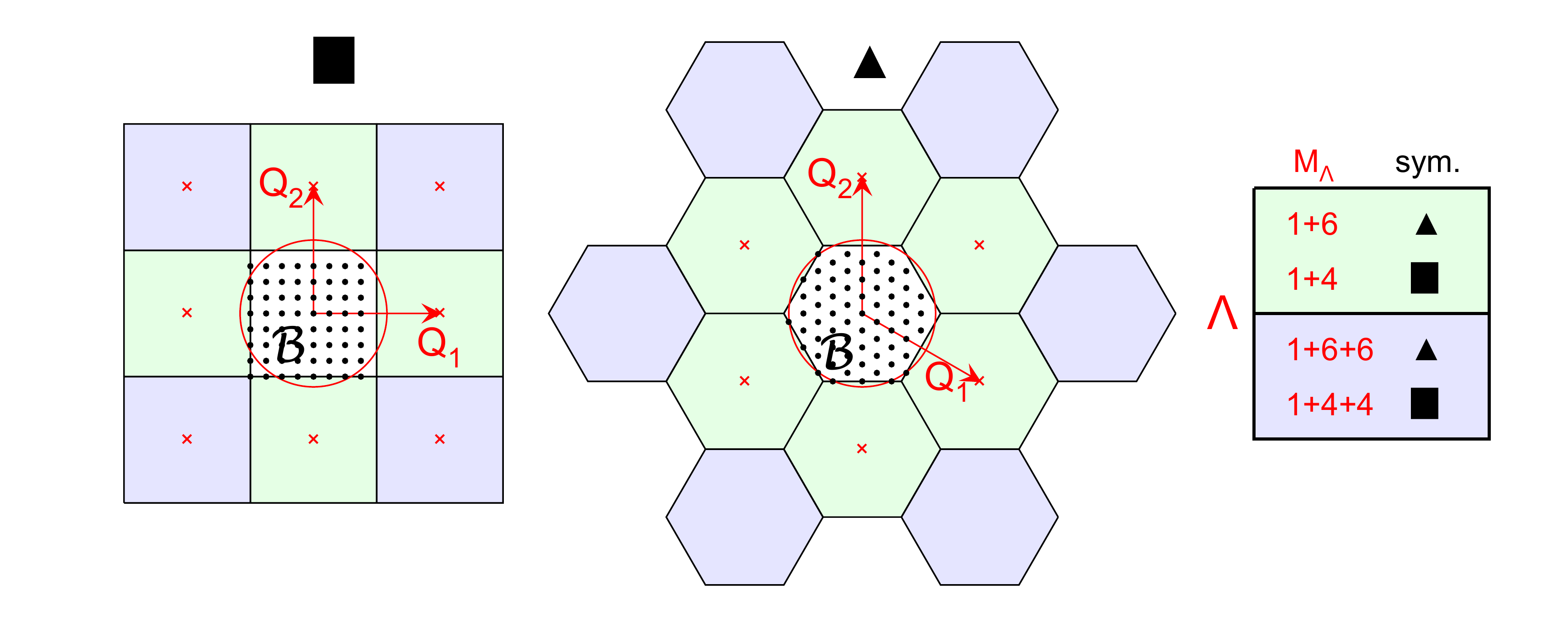}
\caption{(color on line) Illustration of the $k$-space in the square (left) and triangular (right) geometry.
At the center of each figure, are shown the brillouin zone  $\B$ (in white)  and 
the corresponding basis vectors $Q_1$ and $Q_2$.
The first and second shell of neighboring cells,
$\B+n_1Q_1+n_2Q_2$,  are shaded in
light-green and light blue, respectively. 
For square (resp. triangular) symmetry, the integers $n_i$ of the first and second
shell satisfy $n_1^2+n_2^2=1,2$ (resp. $n_1^2+n_2^2-n_1^{}n_2^{}=1,3$).
The corresponding number of cells are summarized by $M_\Lambda$ in the right column.
Most of the results presented in this paper are done including a number of bands, $\NL$, which corresponds to 
two neighboring shells for the square and one for the triangular geometry. In light-red, we indicate
the elementary cell,
$\B_0$,  used in our numerical calculations;
the $\NB$ black dots are an example of the discretization of the Brillouin zone (here $\NB=M^2$ with $M=8$, as explained in Sec.\ref{SEC-CONV}).
The circle indicates the Fermi surface of a Fermi gas.}
\label{FIG-Brillouin}
\end{center}
\end{figure}

We are
considering $N$ electrons in a finite box of volume $V$, of sizes $L_1$ and $L_2$, with periodic boundary conditions, so that
the momentum $k$  belongs to the lattice $L^*$ generated by $L_1^*$ and $L_2^*$ satisfying $L_iL_j^*=2\pi \delta_{ij}$. 

Within the Hartree-Fock approximation, the energy expectation value is minimized with respect to a single skew-symmetric product of $N$ single particle states. Periodic solutions
are special states which can be described by Bloch waves. Let $\Lambda$ be a sub-lattice of $L^*$ generated  by $Q_1$ and $Q_2$. The Brillouin zone is defined as the Voronoi cell of the origin and a periodic state is given by  $|\varphi_k\rangle=\sum_{q\in \Lambda} a_k(q) |k+q\rangle$,  where $k$ belongs to the Brillouin zone, $\B$
(see Fig.\ref{FIG-Brillouin}). 

As a particular case, the Wigner crystal (WC) is obtained by choosing $\Lambda$ such that the Brillouin zone $\B$ contains exactly $N$ states where $N$ is the number of electrons.
Thereafter, the state is built as $\wedge_{k\in \B} |\varphi_k\rangle$. An upper bound of the ground state energy is obtained by minimizing the coefficients $a_k(q)$ of the Bloch functions.
As $r_s$ approaches zero,
the kinetic energy dominates which is minimized by  the Fermi gas (FG) defined by $k\leq k_F$ (see below).  
Such a state cannot be described by the WC.
In ref.~\cite{HF-2008}, the HF energy of the polarized gas has been minimized without imposing periodicity of the solutions;
nevertheless, at intermediate densities,  the HF ground states  are periodic with larger modulations than in the WC
corresponding to less electrons than states in the Brillouin zone.

In this paper we focus on periodic solutions with arbitrary modulations. 
For a given modulation and for fixed choice of $k$-vectors in the Brillouin zone the energy computation is fast enough to tackle millions of  electrons
as every single state is described by very few parameters only.
However, the minimization with respect to the choice of $k$-vectors in the Brillouin zone becomes a complicated combinatorial problem.
This combinatory problem is simplified within the framework of density matrix.

The one body density matrix, $\rho_1$, is a symmetric positive matrix 
such that $\Tr \rho_1=1$. Provided that $\rho_1\leq1/N$, this matrix can be seen as one body density matrix of a state of $N$ electrons.
In the thermodynamic limit, the two body uncorrelated density matrix can be expressed in terms of $\rho_1$ as
\begin{align}
	\rho_2(1,2;1',2') =\rho_1(1;1')\rho_1(2;2')	-\rho_1(1;2')\rho_1(2;1').
\end{align}
The total energy,
a-priori a function of the reduced one and two body density matrices,  
can be expressed entirely as a functional of $\rho_1$.
Explicitly, we obtain for the energy per particle in atomic units
\begin{align}
\label{EQ-defEU}
	E=&\frac12\sum_{k\in L^*,\sigma}k^2\rho_1(k,\sigma;k,\sigma) 
	+\frac1{ r_s^2}
	\sum_{\substack{q,k_1,k_2\in L^*\\ \sigma_1,\sigma_2}}
			v_q\rho_2(k_1,\sigma_1,k_2,\sigma_2;k_1+q,\sigma_1,k_2-q,\sigma_2)
\end{align}
where 
$v_q=1/\|q\|$ for $q \ne 0$, and $v_0=0$. 
For instance, the unpolarized Fermi gas (U-FG) corresponds to $\rho_1(k\sigma,k'\sigma')=\delta_{kk'}\delta_{\sigma\sigma'}\Theta(k_{F,U}-\|k\|)/N$ with $\pi k_{F,U}^2=2\pi^2N/V$;
the resulting energy is $E_{FG}^{U}=1/(2 r_s^2)-8/(3\pi \sqrt{2} r_s)$, and the
fully polarized Fermi gas (P-FG) corresponds to
$\rho_1(k\sigma,k'\sigma')=\delta_{kk'} \delta_{\sigma+}  \delta_{\sigma\sigma'} \Theta(k_{F,P}-\|k\|)/N$ with $\pi k_{F,P}^2=(2\pi)^2N/V$ and energy $E_{FG}^{P}=1/ r_s^2-8/(3\pi r_s)$. In general, without any specification, $k_F$ denotes  
the Fermi wave vector according to the polarization of the corresponding state.

In the following we restrict the density matrix to represent periodic solutions.
The corresponding one-body density matrix can be written as : 
\begin{align}
\label{EQ-defrhok}
\rho_1(1,1')\equiv \rho_1(k+q,\sigma;k+q',\sigma')\equiv\rho_{k}(q,\sigma;q',\sigma')
\end{align}
with $q,q'\in \Lambda$ and $k$ in the Brillouin zone $\B$.
Thus, the density matrix is now described by a family of positive matrices  $\rho_{k}$ such that $\rho_{k}\leq1/N$ and $\sum_{k}\Tr \rho_{k}=1$.

Numerically, we truncate the number of lattice vectors of the sub-lattice $\Lambda$ and include only the first $\NL$ vectors
of smallest norm in the numerical calculations. In the framework of band structure calculations, where the Bloch states are
obtained from an external periodic potential, $\NL$ corresponds to the
number of bands considered. Thus $\rho_{k}$ is a $2\NL\times 2\NL$ matrix and in order to fulfill the condition $\rho_{k}\leq1/N$,  it is more convenient to write:
\begin{align}
\label{EQ-rhokUDU}
\rho_{k}=U^*_{k}D_{k}U_{k}
\end{align}
where $D_{k}$ is a diagonal matrix with $0\leq D_{k}\leq 1/N$ and $U_{k}$ is a unitary matrix.
The potential energy contains a convolution in momentum space calculated using fast Fourier transform (FFT). 
The minimization of the HF energy is done computing  the gradient of the energy with respect to  $U_{k}$ and $D_{k}$.
The only
drawback of the method is to fulfill  the condition $D_{k}\leq1/N$.

The minimization at given density consists 
in the following steps. At first we choose $D_k$ and $U_k$ to start with. Then 
we find the best $U_k$ with a quadratic descent method\cite{HF-2008}. The next step is to try to improve $D_k$ given the gradient of the energy
with respect to $D_k$ and the linear constrains, $0\le D_k\le1/N$ and  $\sum_{k} D_{k}=1$.
The process stops as soon as  $D_k^{(\rm new)}=D_k$. In this case almost every $D_k$ are
$0$ or $1/N$ and the gradient is negative or positive accordingly. Otherwise, we change $D_k$ into  
$(1-\varepsilon)D_k+\varepsilon D_k^{(\rm new)}$ (with a small $\varepsilon$ to ensure 
that $U_k$ follows  $D_k$ {\sl adiabatically}) and we restart the minimization with respect to $U_k$.

In this work, we study the 2DEG for triangular ($\blacktriangle$) and square ($\blacksquare$) symmetries where
$\|Q_1\|=\|Q_2\|=Q$. 
Starting from a state of arbitrary polarization, the minimization always resulted in either an unpolarized (U) or a fully polarized (P) state. 
The Brillouin zone of the Wigner crystal contains exactly $N$ states, so that
 $Q/k_F =Q_W/k_F =\sqrt{2 \pi/\sqrt3} \approx 1.9046$ for the triangular WC (U or P), whereas
$Q/k_F =Q_W/k_F =\sqrt{\pi} \approx 1.7725$ for the square WC (U or P).
Notice that for triangular symmetry the corresponding direct space lattices are quite different: honeycomb lattice for unpolarized  and triangular lattice for polarized states.
The FG can be reached when the Fermi surface is contained inside the
Brillouin zone, that is for  $Q\geq2 k_F $.
Thus, in our simulations, $Q$ varies between $Q_W$ and $2 k_F$.

\begin{figure}
\begin{center}
\includegraphics[width=0.49\textwidth]{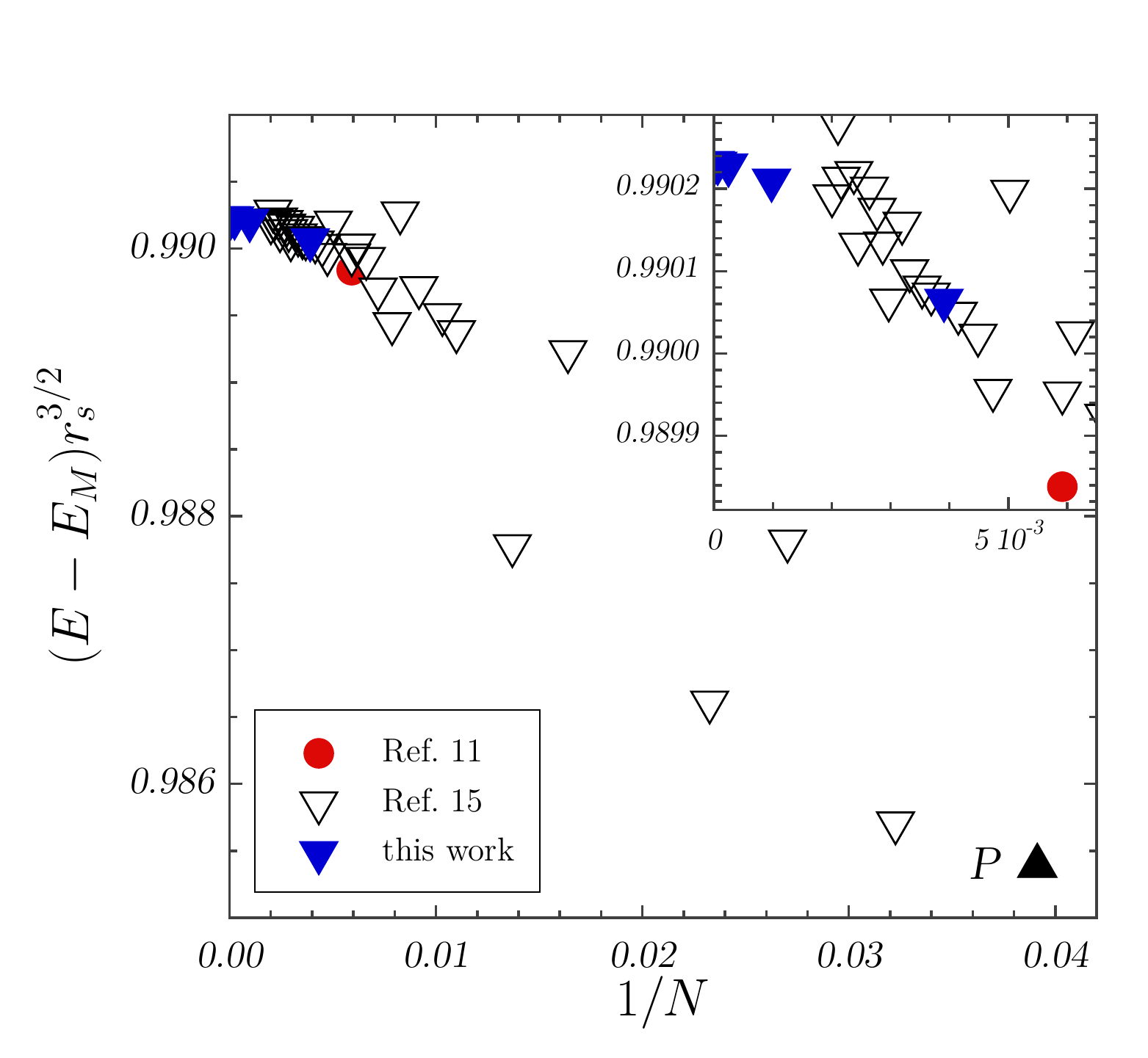}
\includegraphics[width=0.49\textwidth]{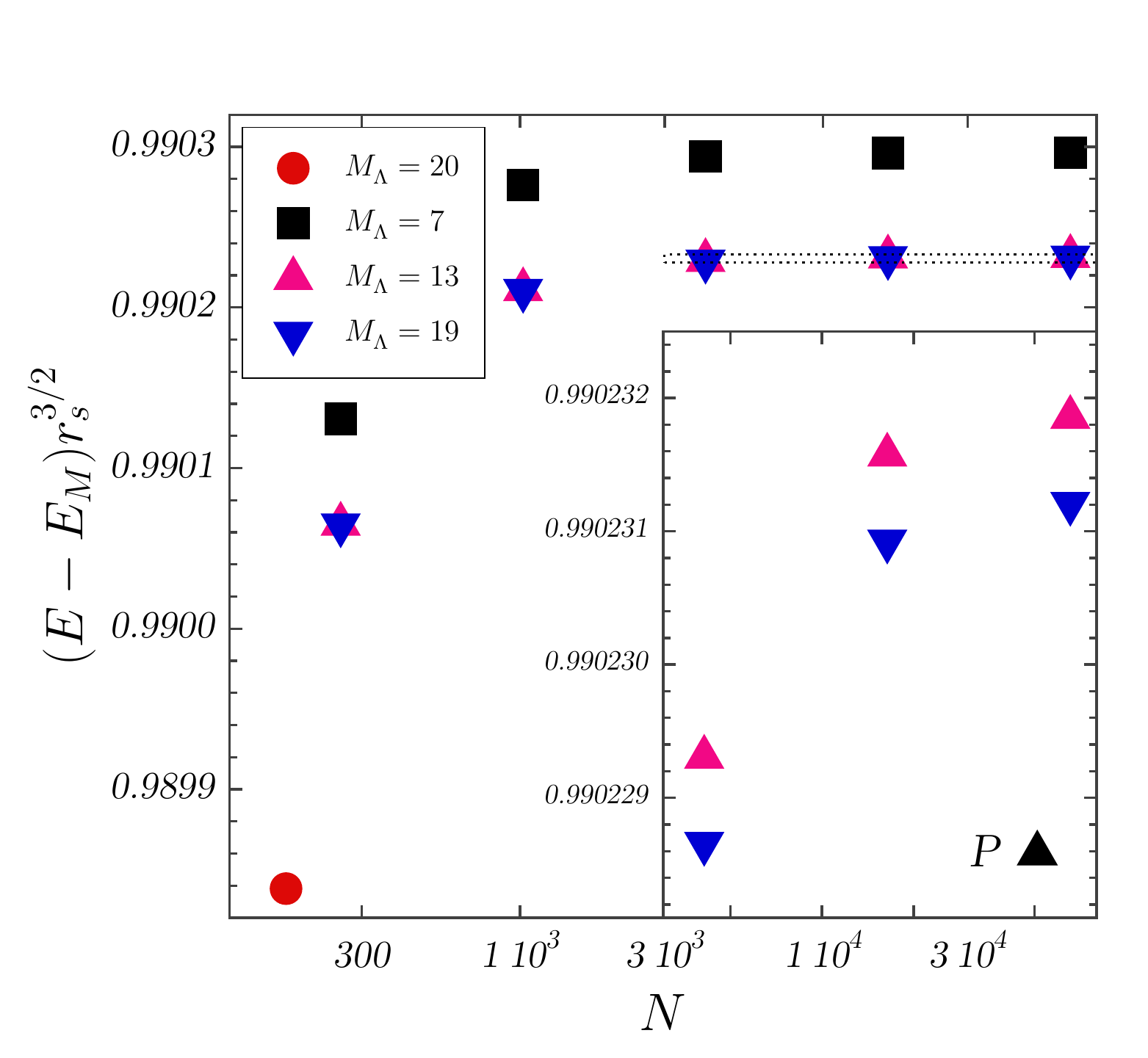}
\caption{(color on line)
Energy (in Hartree units)
 of the 2DEG in the triangular symmetry at  $r_s=4$  (WC)
 as a function of the number of particles, $N\equiv \NB$, and 
the number of included bands, $\NL$.
 $E_M=-1.1061/r_s$ is the Madelung energy.
 ({\bf P}$\blacktriangle$ indicates polarized final state with triangular geometry.)
Left: comparison with previous work\cite{HF-2008,Needs, comparison-Needs}.
Blue full down triangles are results of the present work using $\NL=19$.
Right: convergence with respect to $N$ and $\NL$. 
The inset is a zoom of the dotted-line-domain.
}
\label{FIG-comp}
\end{center}
\end{figure}

\section{Convergence studies}
\label{SEC-CONV}

We first focus on size effects in the thermodynamic limit extrapolation, $N \to \infty$. 
We set $Q_i=ML_i^*$, thus the Brillouin zone contains $\NB=M^2$ vectors.
Since $N/ \NB=(Q_W/Q)^2$, this limit at fixed $Q$ is equivalent to study the convergence with respect to $\NB$.
Fig.~\ref{FIG-comp}  shows the size extrapolation of the 2DEG in the triangular symmetry  at  $r_s=4$ ($Q=Q_W$),
together with the results of Trail et al.\cite{Needs}, done at $\NB=13$ and $\NL\simeq20$,
and those of  Ref.~\cite{HF-2008}. As the calculations of Ref.~\cite{HF-2008} do not assume any periodicity in the
HF search, they are limited to  system sizes $N \lesssim 500$, and the extrapolation to the thermodynamic limit is less accurate.

\begin{figure}
\begin{center}
\includegraphics[width=0.40\textwidth]{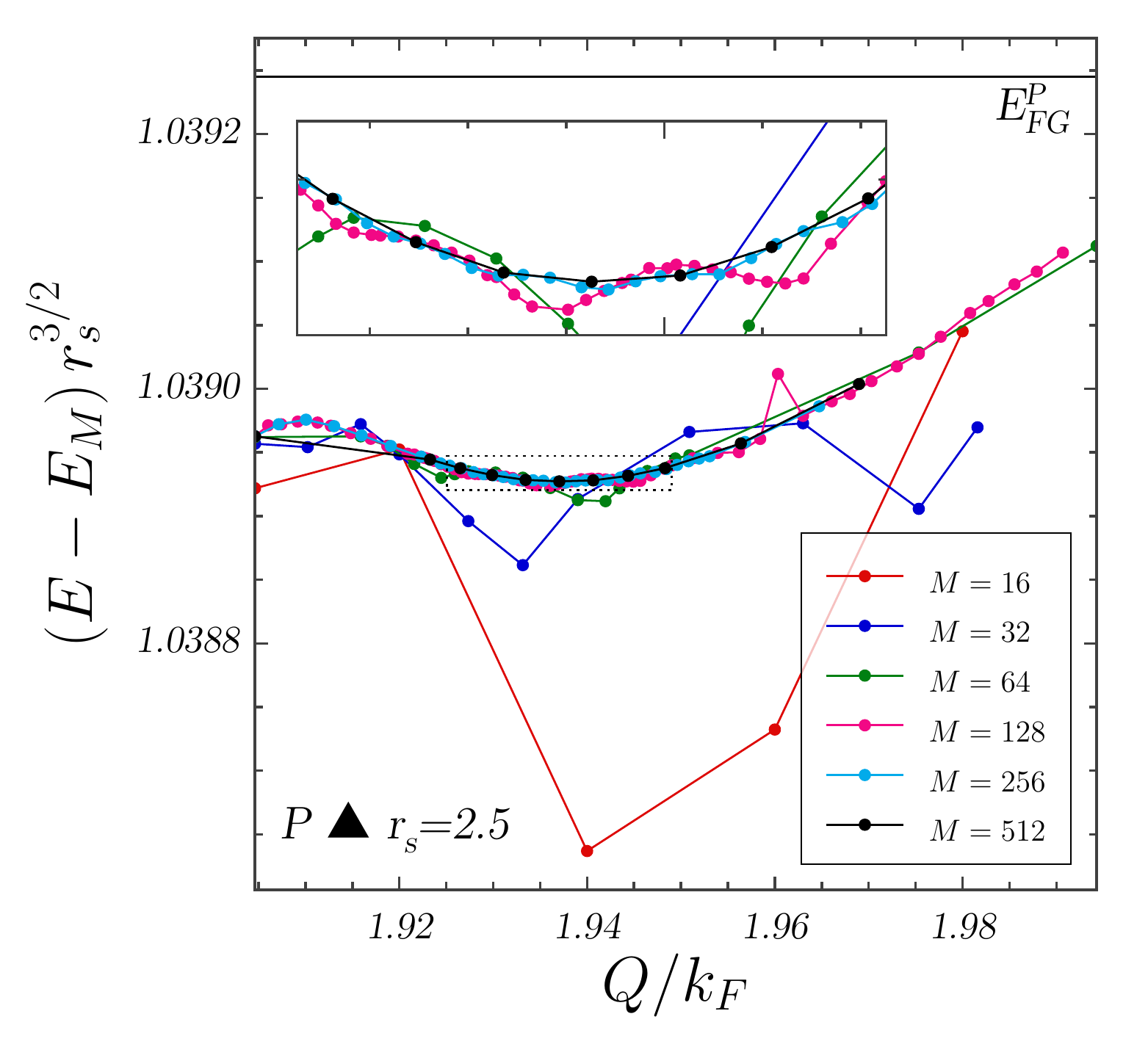}
\caption{
Variation of the energy of the 2DEG at $r_s=2.5$ with triangular symmetry versus $Q$ for different values of $\NB =M^2$.
Inset is the zoom of the region represented by the dashed rectangle.
}
\label{FIG-EQrs24}
\end{center}
\end{figure}

Size effects depend on the phase considered. In the incommensurate phase,
size corrections are not any more monotonic functions, as in the Wigner crystal, but oscillatory behavior
occurs 
depending on the density $r_s$, and on the
modulation vector $Q$.
In Fig.\ref{FIG-EQrs24}, we show the energy of the 2DEG in a triangular symmetry at $r_s=2.5$ versus the modulation $Q$
 (incommensurate crystal) for various system sizes using $M=2^p$, with $p$ from 4 to 9 ($\NB=16^2$ up to $512^2$).
Note the random like oscillations due to the discretization, $\NB$, of the Brillouin zone.
However, at large enough $\NB$, these oscillations are sufficiently small to  analyze safely $E(Q,r_s)$ as seen in Fig.\ref{FIG-HFRES-FITS}.

\begin{figure}
\begin{center}
\includegraphics[width=0.40\textwidth]{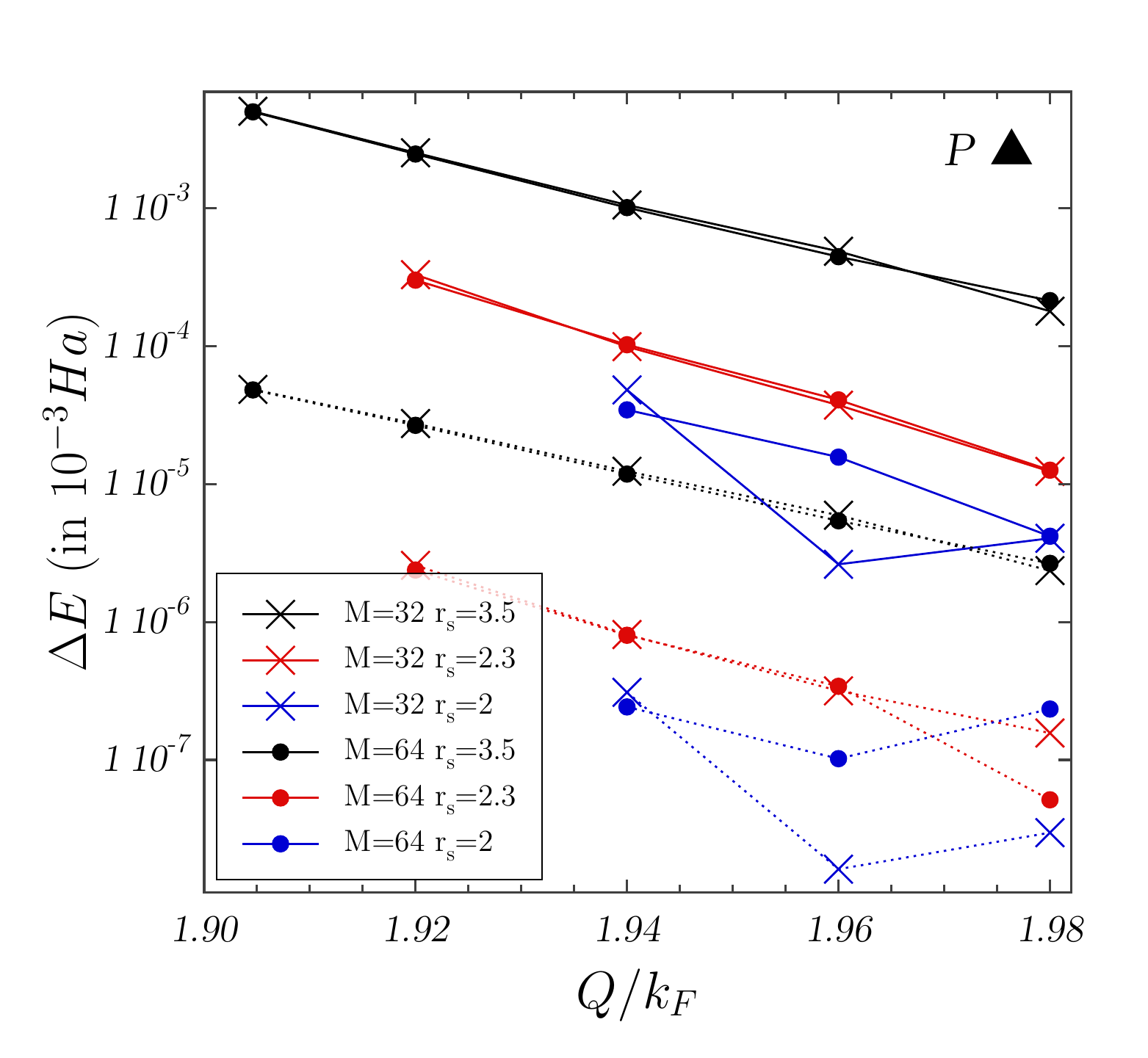}
\caption{(color on line)
Convergence of the energy with respect to $\NL$ for the 2DEG with triangular symmetry for two system sizes $M=32$ and $M=64$.
Full and dotted lines stands $\Delta E=E_{M_{\Lambda,1}}-E_{M_{\Lambda,2}}$ and $\Delta E=E_{M_{\Lambda,2}}-E_{M_{\Lambda,3}}$, respectively,
with  $M_{\Lambda,1}=7$, $M_{\Lambda,2}=13$, $M_{\Lambda,3}=19$.
Crosses and dots stand for $M=32$ and $64$, respectively.
(Values at $r_s=2$ are close to the convergence threshold of the descent method.)
}
\label{FIG-COMPNSH}
\end{center}
\end{figure}

Our second parameter is the number of vectors $\NL$ considered in $\Lambda$.  Note that truncation of $\Lambda$ does not
violate the variational principle, so that the energy of a converged HF solution must decrease as $\NL$ increases.
Figure~\ref{FIG-comp}  and Fig.~\ref{FIG-COMPNSH}  show the convergence in system size $\NB$ (discretization of the Brillouin zone) together with the
exponential convergence in $\NL$ which measures the large $k$ importance. 
As expected, energies decrease with $\NL$ because the Hilbert space is increased.
Interestingly, the $\NL$ improvement is mainly independent of $\NB$ (see Fig.\ref{FIG-comp}-right and Fig.\ref{FIG-COMPNSH}), which allows us to work with small $\NL$ and 
estimate corrections using small systems. 
Most of the calculations presented in this paper are thus performed with $\NL=7$ and $\NL=9$
bands for supercells of triangular and square symmetry, respectively.

\section{Results}
\label{SEC-RES} 

We have studied the HF ground state of the 2DEG in the density region $0.8 \le r_s \le 4$ at zero temperature
considering commensurate and incommensurate solutions with square and triangular symmetries.
At low densities the electrons form a commensurate Wigner crystal of modulations $Q=Q_W$ and we recover the results of previous HF studies
\cite{Needs,HF-2008,comparison-Needs}. For higher densities, an incommensurate crystal with modulation  $Q_W < Q  < 2k_F$ is formed 
for any fixed polarization and symmetry. 
\begin{figure}
\begin{center}
\includegraphics[width=0.32\textwidth]{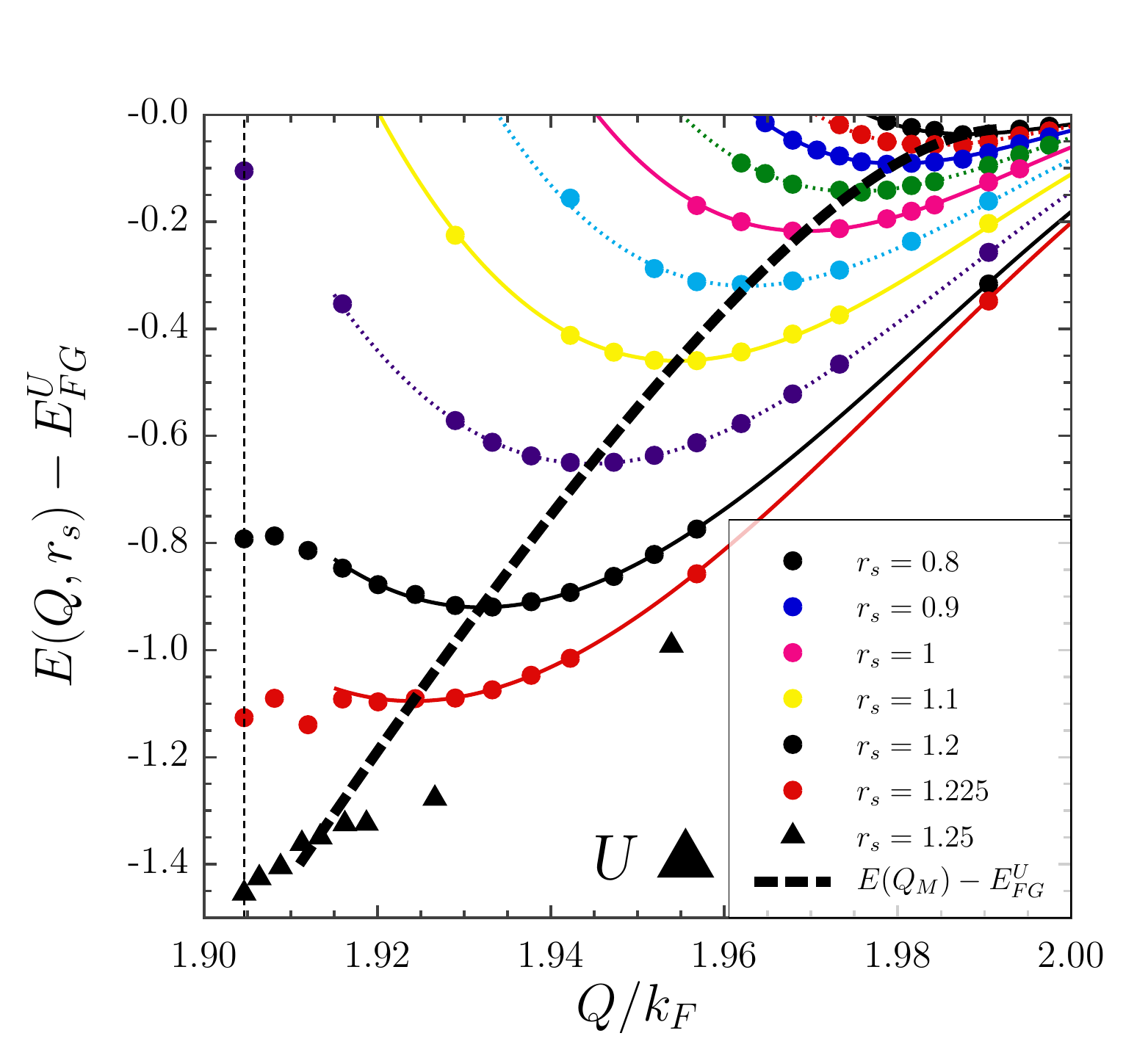}
\includegraphics[width=0.32\textwidth]{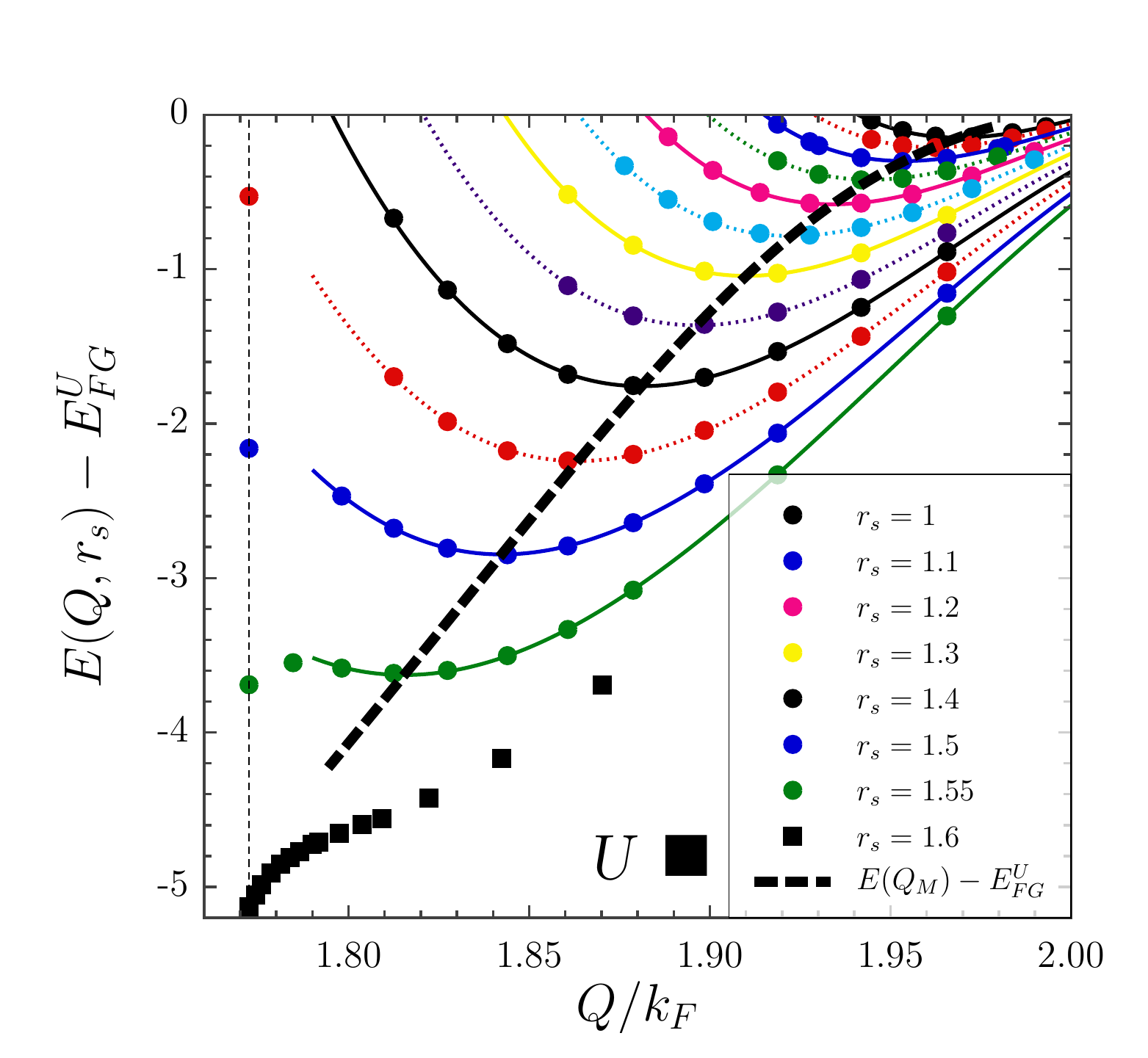}
\includegraphics[width=0.32\textwidth]{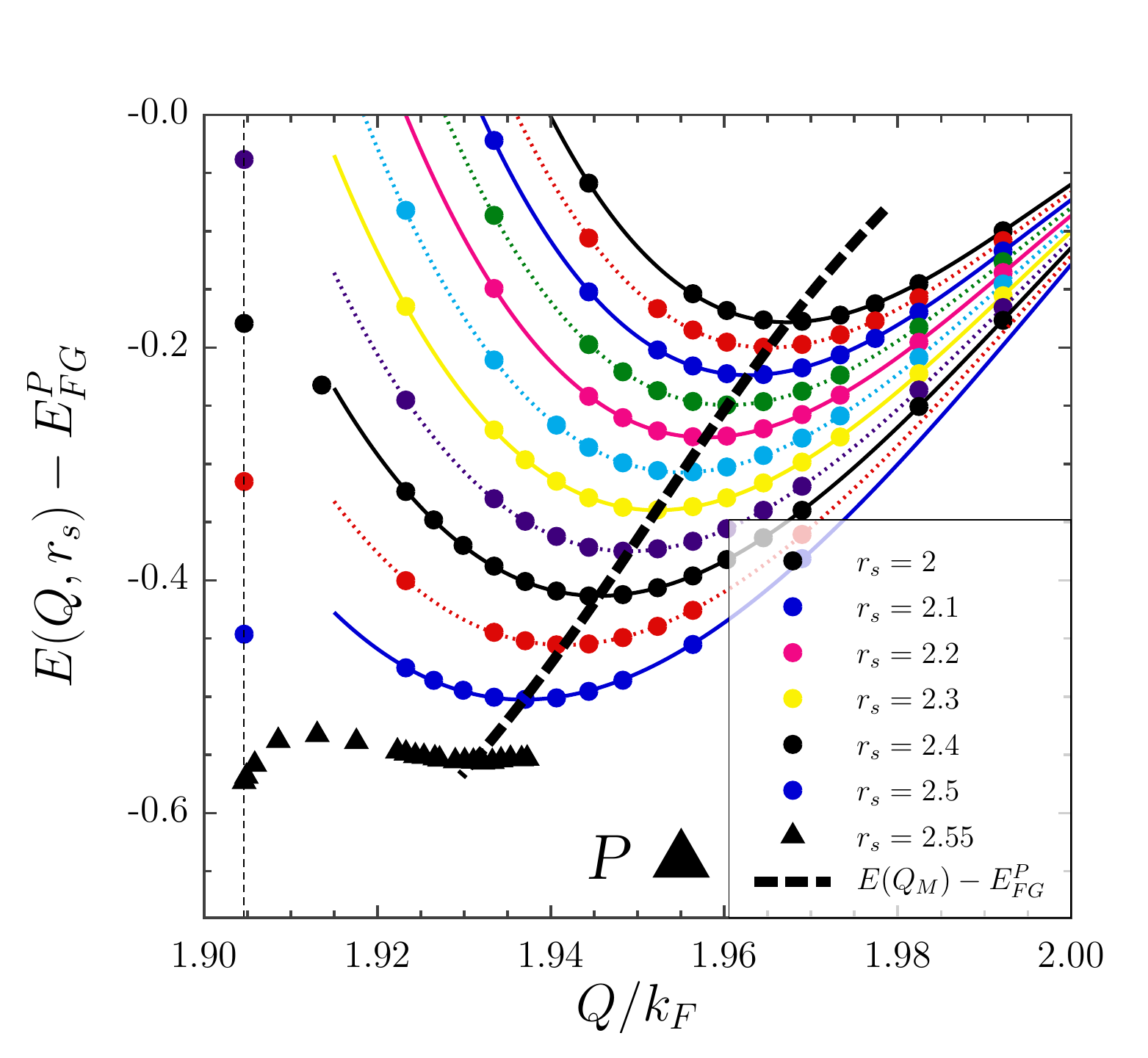}
\caption{
(color on line) 
Energy difference, with respect to the Fermi gas, $E(Q,r_s)-E_{FG}^{U/P}$,  in milli Hartree versus 
modulation, $Q$,
for different densities  and symmetries ($\blacktriangle$ or $\blacksquare$) at $\NB=256\times256$.
The final polarization obtained after minimization is either unpolarized (U) or fully polarized (P).
Lines are the polynomial fits using the parameters given in Table-\ref{TABLE-COEFFS}.
In each figure, the lowest curve (largest $r_s$) with triangular or square symbols has a minimum at $Q=Q_W$.
The bold dashed-line connects $Q_M(r_s)$, the minima of $E(Q,r_s)$ for fixed $r_s$.
Vertical dotted lines indicate $Q_W$.}
\label{FIG-HFRES-FITS}
\end{center}
\end{figure}

\begin{figure}
\begin{center}
\includegraphics[width=0.45\textwidth]{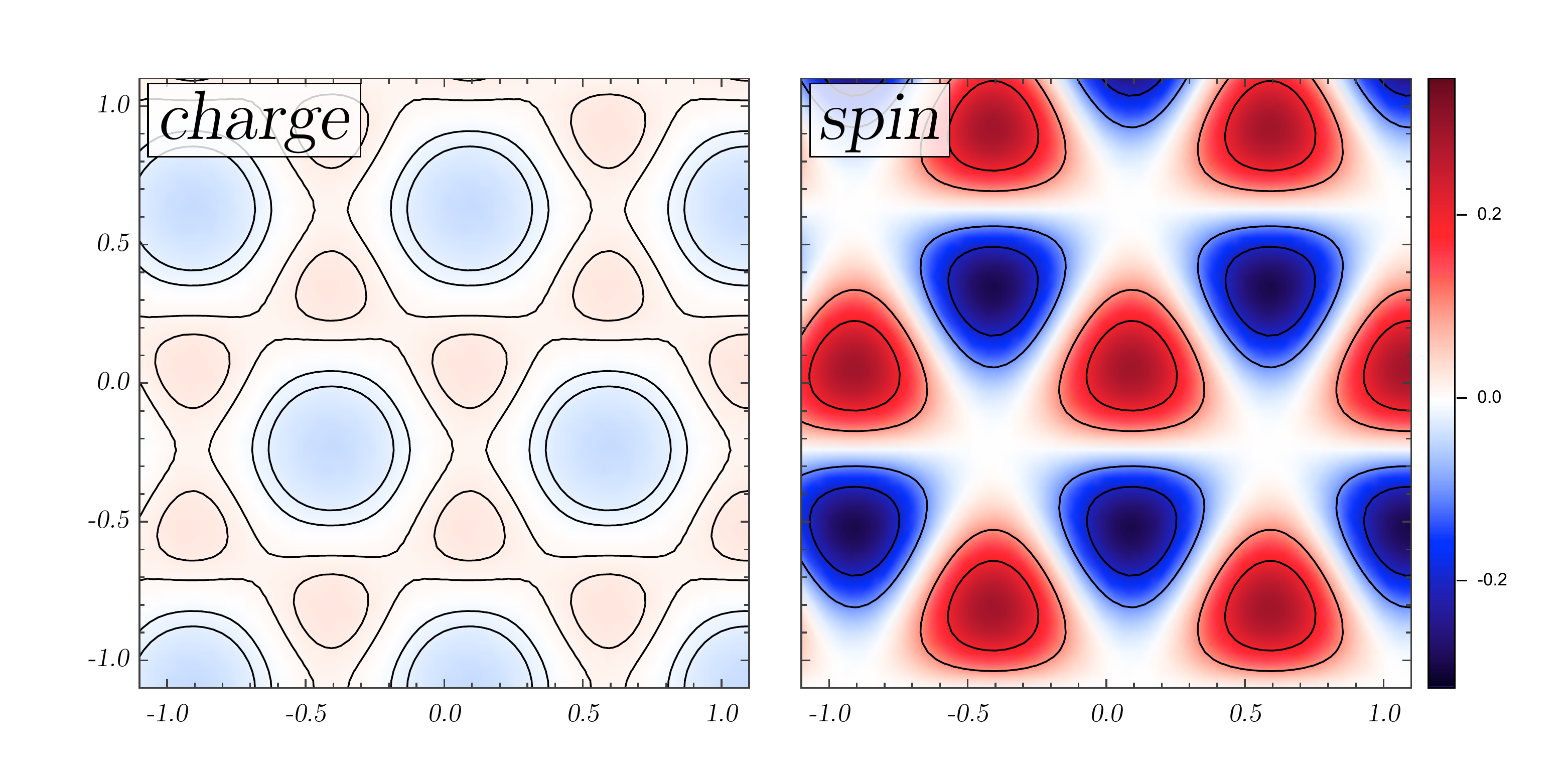}
\includegraphics[width=0.45\textwidth]{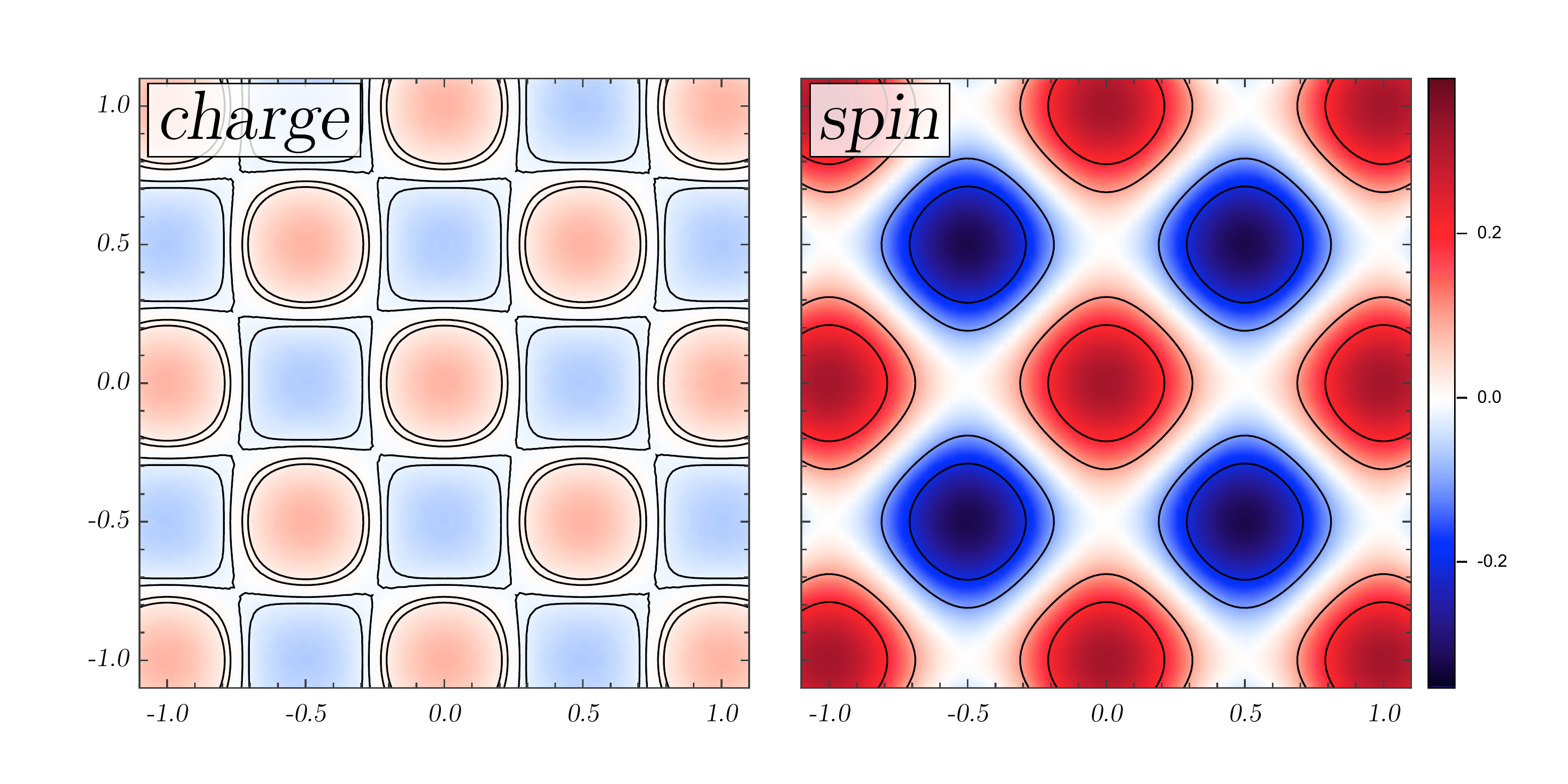}
\caption{(color on line) One body charge and spin densities of an unpolarized incommensurate crystal with triangular symmetry  (left: $r_s=1.2$, $Q/k_F=1.933$, $N/\NB\simeq0.97$) and square symmetry (right: $r_s=1.5$, $Q/k_F=1.844$, $N/\NB\simeq0.92$).
Average values have been subtracted. Lengths are given in units of the inverse modulation, $Q^{-1}$.
The color scaling is the same for all pictures. 
Contour levels are 
$\pm0.01$, $\pm0.02$ for the charge densities
and 
at $\pm0.1$, $\pm0.2$ for the spin densities .}

\label{FIG-RHOx}
\end{center}
\end{figure} 
Figure \ref{FIG-HFRES-FITS}  summarizes the energy gain with respect to the unmodulated Fermi gas
as a function of $Q$ at different densities.
Well inside the incommensurate phase ($Q>Q_W$),  the energies can be well represented 
with a polynomial form:
\begin{eqnarray}
\label{EQ-fitdef}
	E(Q,r_s)&=&E_{FG}(r_s)+\sum_{i=0}^3 \sum_{j=0}^2 \alpha_{ij} X^i r_s^j
\end{eqnarray}
where $X=100(Q/k_F-2)$.
The parameters $\alpha_{ij}$ determined by least square fits are given in Table-\ref{TABLE-COEFFS}. 
From this parametrization, for fixed $r_s$,
we determine the minimum $ Q_M(r_s)$ of $E(Q,r_s)$, shown in Fig.\ref{FIG-HFRES-FITS}.

\begin{figure}
\begin{center}
\includegraphics[width=0.99\textwidth]{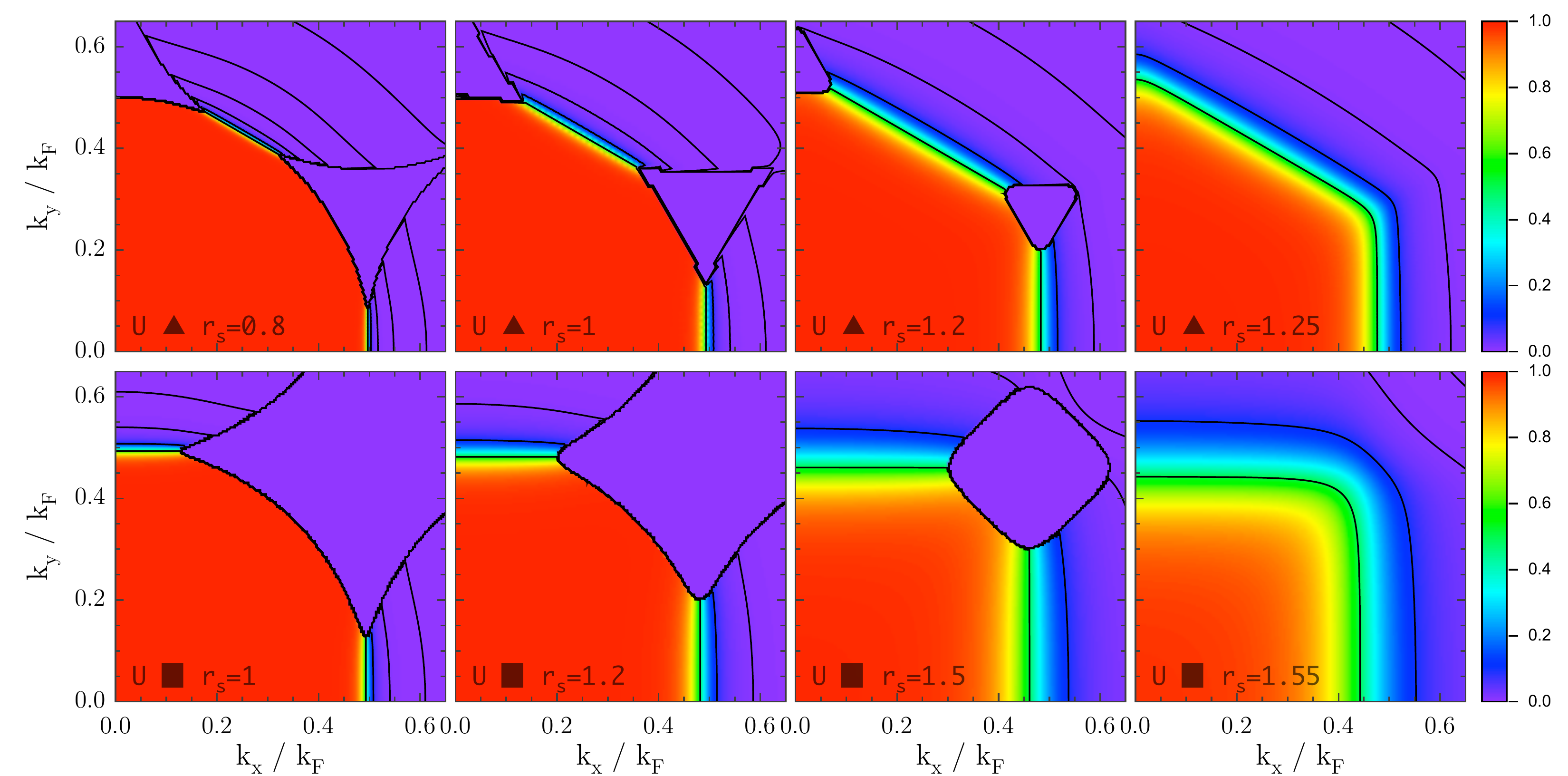}
\caption{(color on line) Modulus of the momentum distribution $|n_\uparrow(k)|=|n_\downarrow(k)|$ of U solutions
as a function of wave vector $k$ in the positive quadrant (other parts can be deduced by symmetry).
Top line and bottom line are for triangular and square symmetry, respectively.
Contour levels are at 0.5 and  0.1, 0.01, etc.
From right to left is shown the evolution from the Wigner crystal distribution (continuous function everywhere) to Fermi gas with a step along some directions. 
For the incommensurate states, note the step-function behavior to a domain where $n_k=0$ which grows in the corner of the Brillouin zone when $r_s$ decreases.
}
\label{nk-all}
\end{center}
\end{figure}

The incommensurate phase is characterized by a crystal in direct space with slightly more lattice sites $\NB$
than electrons $N$, increasing for larger modulation according to $\NB/N=(Q/Q_W)^2$.
Figure \ref{FIG-RHOx} shows typical charge and spin densities in the incommensurate phase for the triangular and square geometry.
The two examples are chosen close to the transition to the {\sl Wigner} crystallization.
The amplitude of the modulation of the charge densities is about an order of magnitude smaller than that of the spin densities, an
effect which is even more pronounced at higher density.

The momentum distribution $n_k$ ($N$ times the diagonal part of $\rho_1$)
provides additional insight. In contrast to the step-function behavior at $k_F$ of the Fermi gas, 
$n_k$ is continuous inside the
commensurate Wigner crystal phase and its variation reflects the
symmetry of the Brillouin zone. The incommensurate phase still reflects the underlying symmetry of the crystal, but 
angle selective steps occur at the corners of the Brillouin zone (see Fig. \ref{nk-all}). The rounding of the corners increases 
for smaller $r_s$, and the isotropic step-function of the Fermi gas is continuously approached for $r_s \to 0$.

\begin{figure}
\begin{center}
\includegraphics[width=0.49\textwidth]{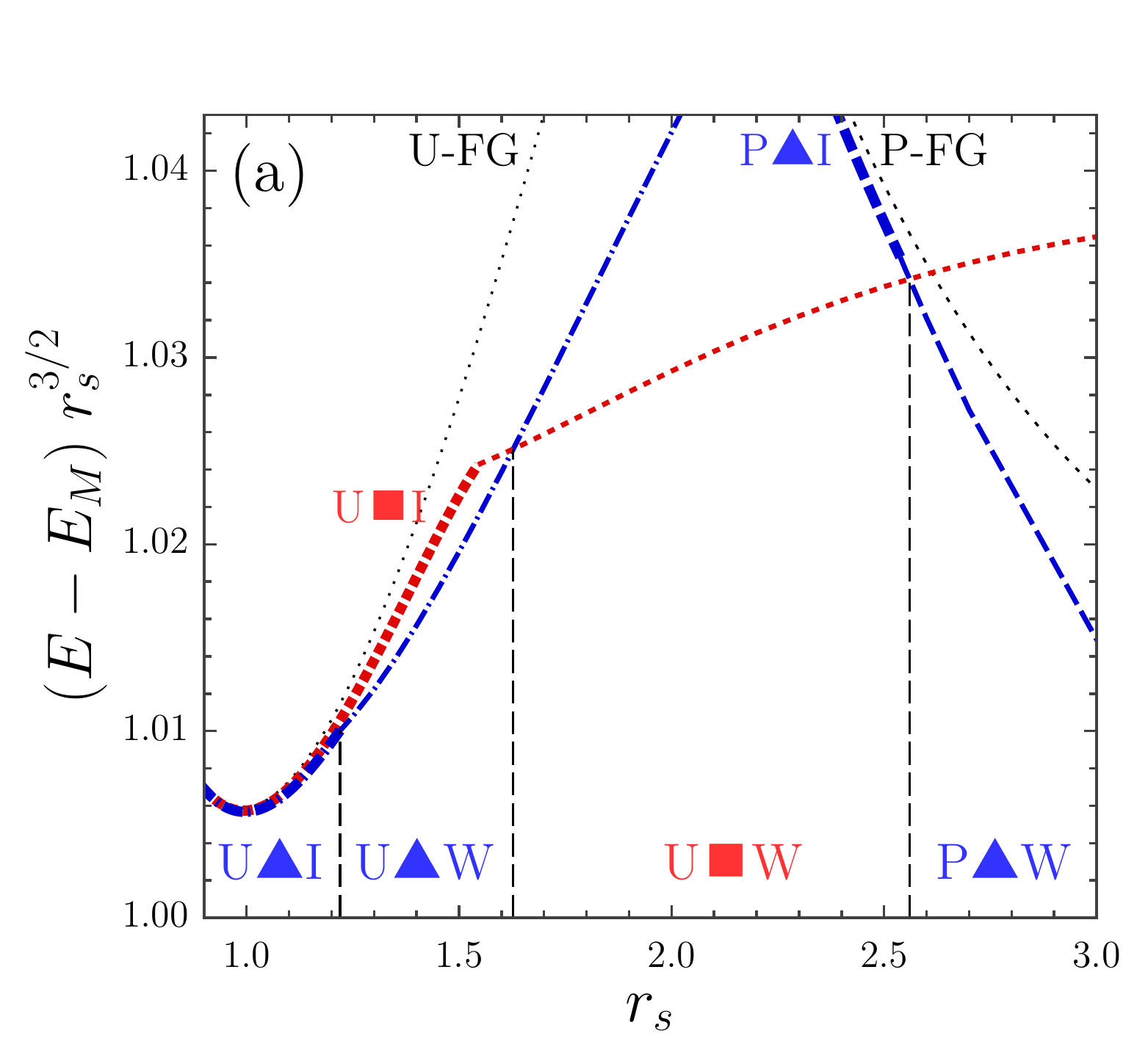}
\includegraphics[width=0.49\textwidth]{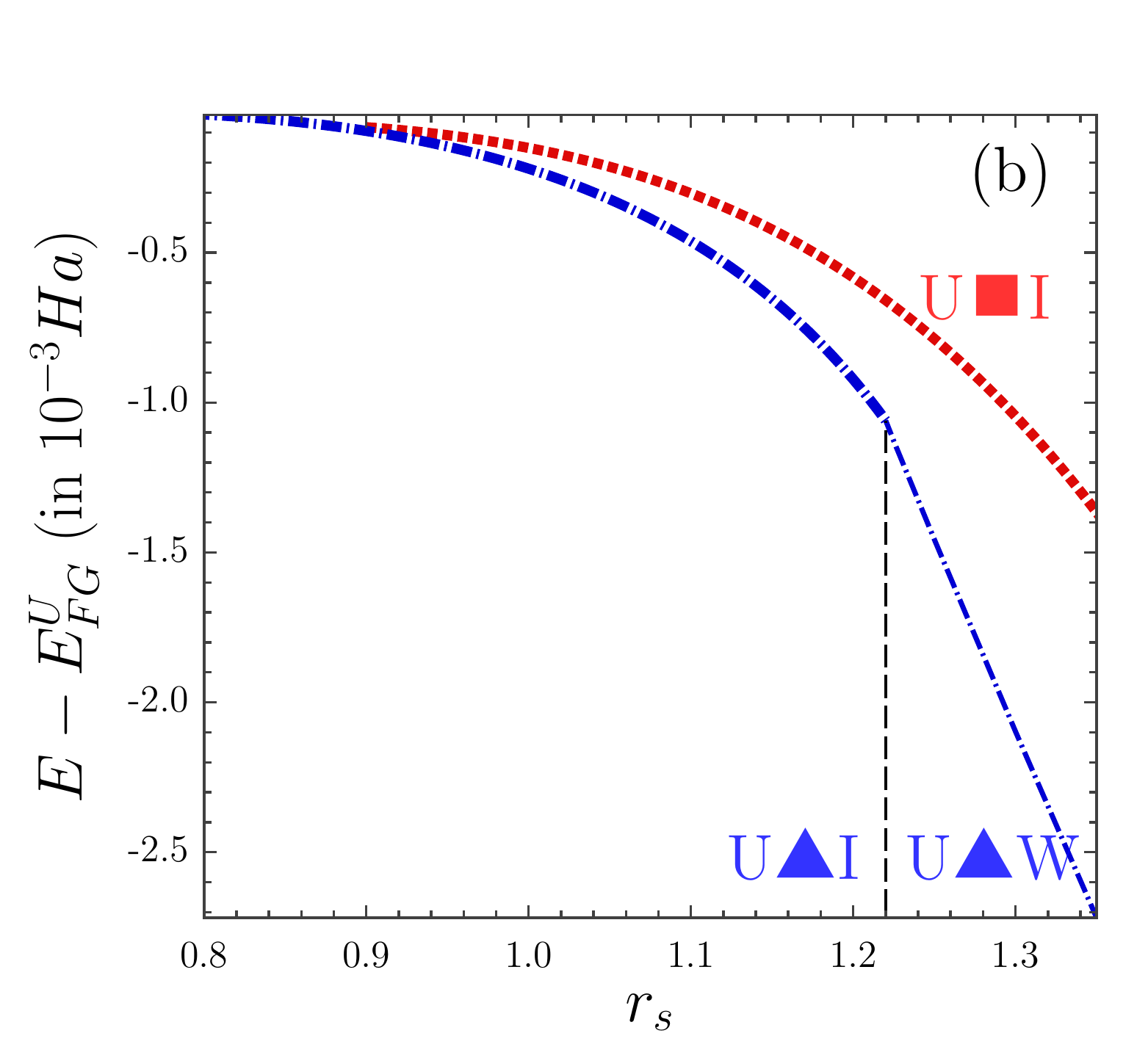}
\caption{
(color online).
$(a)$:
Phase Diagram of the 2DEG at $T=0$, 
where $E_M=-1.1061/r_s$ is the Madelung energy and energies are multiplied by $r_s^{3/2}$.
Dotted lines correspond to the Fermi gas.
Blue and red curves represent the triangular ($\blacktriangle$) and square 
($\blacksquare$) phases, and
$P$ and $U$ stands for polarized and unpolarized phases, respectively, whereas
W and I (thick curves) indicate Wigner crystal and incommensurate crystal, respectively.
The dashed vertical lines indicate the transitions.$(b)$: Energy gain with respect
to the unpolarized Fermi gas energy, $E_{FG}^U$, in the high density 
region.
}
\label{FIG-PhaseDiag}
\end{center}
\end{figure}

Whereas 
we have found that
the incommensurate phase is always favored compared to the Fermi gas solution, independently of the imposed polarization
and crystal symmetry, the unpolarized incommensurate hexagonal crystal
becomes the true HF ground state at high densities, $r_s \le r_s^c \simeq 1.22$. The different phases and energies
for $0.8 \le r_s \le 4.0$ are illustrated in Fig.\ref{FIG-PhaseDiag}. Although our HF method does not impose the polarization,
we have not found any stable partially polarized ground states. 
At $r_s>r_s^c$ the unpolarized electrons form a commensurate Wigner crystal
of hexagonal symmetry, and, at $r_s\simeq 1.62$ a
structural transition from  the  unpolarized hexagonal WC to the unpolarized square WC occurs, followed by a 
transition from the unpolarized square WC to the fully polarized triangular WC at $r_s \simeq 2.6$.

\section{Conclusion}
\label{SEC-Conclusions}

We have studied the 2DEG in the Hartree-Fock approximation at densities $r_s \lesssim 4$.
We
confirm previous observations of incommensurate phases of the fully polarized electron gas\cite{HF-2008}, 
performing calculations of much larger system sizes.
We further included electron polarization, as well  as square and triangular
symmetries.
Our HF phase diagram at zero temperature is much richer than that obtained previously \cite{Needs},
which did not consider the unpolarized triangular WC, nor any incommensurate phase.
Our numerical calculations
explicitly confirm the old conjecture of Overhauser \cite{Overhauser, Giulani}  that Fermi gas is never the HF ground state
which has been proven rigorously for the fully polarized electron gas \cite{HF-2008}.

We have further shown that the momentum distribution provides an unambiguous characterization of
the incommensurate phase.
In contrast to the isotropic momentum distribution of 
a Fermi liquid, discontinuous at  the Fermi surface\cite{momk2D,momk3D},
 the incommensurate phase exhibits an anisotropic momentum distribution intermediate between a crystal and the Fermi gas with  forbidden domains inside the Brillouin zone, where $n_k$ jumps to zero.

\begin{table}
\caption{Coefficients $\alpha_{ij}$ of the polynomial fits $E(Q,r_s)-E_{FG}(r_s)$ defined by Eq.\ref{EQ-fitdef}.}
\begin{center}
\begin{tabular}{rrr|rrr|rrr}
\multicolumn{3}{c|}{U-$\blacktriangle$}
&\multicolumn{3}{c|}{U-$\blacksquare$}&
\multicolumn{3}{c}{P-$\blacktriangle$}\\
\hline
      -0.78611  &        1.88240  &       -1.13180
&       -0.621900  &        1.53040  &      -0.96941
&      0.15758  &     -0.08577  &     -0.011495 
 \\
       0.35614  &      -0.64435  &        0.34780
&      0.058858  &      -0.26652  &       0.27359
&       0.24875  &      -0.24028  &       0.070520
  \\ 
       0.10624  &      -0.15804  &      0.05531 
&      0.032321  &     -0.06425  &      0.02500
&      0.11155  &      -0.10668  &      0.024822
  \\ 
    -0.00166  &   -0.00044  &    0.00081 
&     -0.022986  &      0.02665  &    -0.00750 
&  -0.00090  &    -0.00321  &     0.001310  
\end{tabular}
\end{center}
\label{TABLE-COEFFS}
\end{table}


\begin{thebibliography}{99}

\bibitem{Na} S. Huotari, J. A. Soininen, T. Pylkk{\"a}nen, K. H{\"a}m{\"a}l{\"a}inen, A. Issolah, A. Titov, J. McMinis, J. Kim, K. Esler, D. M. Ceperley, M. Holzmann, and V. Olevano, Phys. Rev. Lett. {\bf 105}, 086403 (2010).

\bibitem{quasi2D} B. Bernu, F. Delyon, and M. Holzmann, 
Phys. Rev. B {\bf 82}, 245116 (2010).

\bibitem{hetero} T. Ando, A.B. Fowler, and F. Stern, Rev. Mod. Phys. {\bf 54}, 437 (1982).


\bibitem{Wigner} E. P. Wigner, Trans. Faraday Soc. {\bf 34}, 678 (1938); Phys. Rev. {\bf 46}, 1002 (1934).

\bibitem{QMC3D} D. M. Ceperley and B. J. Alder, Phys. Rev. Lett. {\bf 45}, 566-569 (1980).

\bibitem{QMC2D} B. Tanatar and D.M. Ceperley, Phys. Rev. B {\bf 39}, 5005 (1989).

\bibitem{exchange} B. Bernu, L. C{\^a}ndido, D. Ceperley Phys. Rev. Lett. 86, 870 (2001).

\bibitem{exchange3D} L. C{\^a}ndido, B. Bernu, and D.M. Ceperley, Phys. Rev. {\bf B 70}, 094413 (2004). 


\bibitem{Overhauser} A. W. Overhauser, Phys. Rev. Lett. {\bf 4}, 462 (1960); Phys. Rev. {\bf 128}, 1437 (1962).

\bibitem{Giulani} G. F. Giuliani and G. Vignale, {\it Quantum Theory of the Electron Liquid}, Cambridge University Press, Cambridge (2005).

\bibitem{HF-2008} B. Bernu, F. Delyon, M. Duneau, and M. Holzmann, Phys. Rev. B 78, 245110 (2008); cond-mat/0810.3559.

\bibitem{Spivac} B. Spivak and S.A. Kivelson, Phys. Rev. {\bf B 70}, 155114 (2004).

\bibitem{Hexatic} B. K. Clark, M. Casula, and D. M. Ceperley, Phys. Rev. Lett. {\bf 103}, 055701 (2009). 

\bibitem{Waintal} H. Falakshahi and X. Waintal, Phys. Rev. Lett. {\bf 94}, 046801 (2005), X. Waintal, Phys. Rev.  {\bf B 73}, 075417  (2006).  

\bibitem{Needs} J. R. Trail, M. D. Towler, and R. J. Needs, Phys. Rev. {\bf B 68}, 045107 (2003).

\bibitem{Shiwei} S. Zhang and D. M. Ceperley, Phys. Rev. Lett. {\bf 100}, 236404 (2008). 


\bibitem{comparison-Needs}
Using the same parameters, we recover exactly the results of Trail et al.\cite{Needs} for the 2DEG  (P$\blacktriangle$) at $r_s>3$. We thank the authors for sending us their data.

\bibitem{momk2D} M. Holzmann, B. Bernu, V. Olevano, R. M. Martin, and D. M. Ceperley, Phys. Rev. B {\bf 79}, 041308 (2009).

\bibitem{momk3D} M. Holzmann, B. Bernu, C. Pierleoni, J.
McMinis, D.M. Ceperley, V. Olevano, and L. Delle Site, arXiv:1105.2338 (2011).
\end{thebibliography}
\end{document}